\documentclass[reprint,amsmath,amssymb,aps,prb]{revtex4-2}

\usepackage{graphicx}
\usepackage{bm}
\usepackage{hyperref}
\usepackage{xcolor}
\usepackage{comment}
\usepackage{fancyhdr}
\RequirePackage{qtree}
\RequirePackage{braket}
\RequirePackage{tikz}

\usepackage{amsmath}
\usepackage{setspace}
\usepackage{fancyhdr}
\usepackage{lastpage}
\usepackage{graphicx}
\usepackage{amssymb}
\usepackage{xcolor}
\usepackage{soul}

\renewcommand{\vec}[1]{\mathbf{#1}}

\begin{document}

\title{Spin singlets are useful}

\author{Silas Hoffman}
\author{Edward H. Chen}
\author{Matthew Brooks}
\author{Charles Tahan}\affiliation{Microsoft Discovery \& Quantum, One Microsoft Way, Redmond, WA 98052, USA}
\author{Stephen Carr}
\author{Daniel Volya}
\author{Alan Tran}
\author{Tyler Keating}
\author{Thaddeus D. Ladd}%
\affiliation{HRL Laboratories, 3011 Malibu Canyon Rd., Malibu, CA 90265, USA}

\date{\today}

\begin{abstract}
We evaluate the utility of the spin-zero manifold of an exchange-coupled array of $N$ spins for tasks in quantum computation and quantum simulation.
Since pairs of electrons can be readily initialized into a product state of singlets in semiconducting quantum dot arrays, the full spin-zero manifold is available with exchange-only control, providing a Hilbert space of approximate dimension $2^N/(N/2)^{3/2}$, asymptotically close to the $2^N$ dimension of the full spin Hilbert space.
Leveraging the spin-zero manifold enables larger computational space in a given array compared to traditional exchange-only control, in which spin arrays are organized into modular units of $n$ spins comprising $N/n$ encoded qubits, limiting to the exponentially smaller Hilbert dimension $2^{N/n}$.
Here we focus on benchmarking metrics for this resource utilization by generalizing cross-entropy benchmarking, mirror benchmarking, and out-of-time-ordered correlators to this system.   
We show that operating in the spin-zero manifold can accelerate the realization of computational quantum advantage applications in semiconductor-based spin qubits.
\end{abstract}

\maketitle

\section{Introduction} 
Fault-tolerant quantum computing demands physical qubits that combine low error rates with a path to large-scale integration. 
In semiconductor spin qubits, one promising approach is the exchange-only (EO) qubit, {where three spins encode a single qubit of information in a decoherence-free subsystem}~\cite{%
	bacon_universal_2000,%
	DiVincenzo2000,%
	kempe_theory_2001,%
	medford_self-consistent_2013,%
	Eng2015,%
	weinsteinNAT23}.  
The name ``exchange-only'' reflects the use of the Heisenberg \textit{exchange} interaction to achieve universal qubit control using purely baseband electrical signals, thereby eliminating the need for microwave drives or magnetic field gradients~\cite{DiVincenzo2000}. 
EO qubits are also intrinsically immune to global magnetic fluctuations \cite{kempe_theory_2001}. 
However, this encoding has two major deficiencies: (1) three spins for one qubit is a large overhead, i.e. an exponentially vanishing fraction of the total Hilbert space is utilized, and (2) two-qubit gates require tens of exchange pulses which lower encoded gate fidelity and traverse the noncomputational elements of the Hilbert space, leading to leakage from the subsystem under exchange error~\cite{DiVincenzo2000,Fong2011,HRL2qubit2022}. 
Many other types of encodings have been proposed as well, all designed to overcome the limitations of single-spin systems \cite{taylor_electrically_2013,shimPRB16,salaPRB17,foulkPRL25,boscoPRA26}.
Despite these near-term challenges, encoding offers the potential for robust universal fault tolerant quantum computing.

Here, we propose an alternative perspective that abandons modular qubit encoding entirely.
Initializing $2L$ spins as $L$ singlet pairs creates a state within the total-spin-zero (S0) manifold.
S0 refers to the manifold of total spin-states whose action under global SU(2) rotations correspond to the trivial irreducible representation---that is, they are the degenerate eigenstates of any global rotation. 
The Heisenberg exchange interaction between any pair of spins is an operator that also lies in this trivial irreducible representation.  
Algebraically, projections into S0 states and exchange operators all commute with the total spin operator $\vec{S}=\sum_{\ell=1}^{2L} \vec{S}_\ell$, i.e. the sum of all commuting single-spin vector operators $\vec{S}_\ell=\{S_\ell^x,S_\ell^y,S_\ell^z\}.$
As such, any exchange pulse sequence, even if distorted by noise, preserves the S0 manifold exactly.  
Hence there is no leakage by construction, unless caused by spurious local gradients of magnetic field (which arise, in practice, from hyperfine interactions with nuclear spins, amongst other sources.)
Pairwise exchange-only operation, either with a continuously controllable exchange pulses or with a discrete set of exchange-generated unitaries (for example, only square-root-of-swap operations) offer universal control~\cite{baconPRL00,kempe_theory_2001}.

The S0 subspace dimension grows exponentially with $L$; it is the combinatorially increasing number of ways of grouping an ordered list of the spins.   
This is the $L$-th Catalan number,
\begin{equation}
  C_L = \frac{1}{L+1}\begin{pmatrix}2L \\ L\end{pmatrix}  \approx \frac{4^L}{L^{3/2}\sqrt{\pi}},
  \label{eq:catalan}
\end{equation}
where the latter approximation results from the Stirling approximation, increasingly accurate at large $L$. 
This dimension nearly saturates the full Hilbert space. For example, fifty-four quantum dots~\cite{HRLmultirail2025} yield $L = 27$ singlet pairs spanning $C_{27} > 10^{14}$ states---more than $47$ effective qubits, which is competitive with many leading superconducting processors~\cite{abrahamCM26}.

There are two costs to utilizing the S0 manifold for a spin-qubit array, relative to dividing the array into groups of $n$ spins and using the $2^{N/n}$-dimensional tensor-product space of those groups (typically with $n=3$ or $n=4$ for exchange-only qubits). 
The first is the challenge of constructing methods for fault-tolerant quantum error correction under local error models, since local spin and spin-pair errors are challenging to localize and correct when computing across the manifold of highly entangled S0 states. 
In the present work, we set this challenge aside, either under the presumption that a system of interest has a nonlocal error model (for example, global magnetic field fluctuations to which the whole manifold is immune) or that future, alternative methods of fault-tolerance error resilience emerge.  

The present work focuses instead on a second, related cost, which is the resolution of quantum measurement. 
Exchange-only qubit operation typically employs Pauli spin blockade (PSB), which distinguishes and projects spins into singlet or triplet states. 
For two spins $a$ and $b$ with total angular spin angular momentum $\vec{S}_{ab}=\vec{S}_a+\vec{S}_b$ and $\vec{S}_{ab}\cdot\vec{S}_{ab}=S_{ab}(S_{ab}+1)$, PSB distinguishes \emph{only} the quantum numbers $S_{ab}$ as 0 (called singlet) or 1 (called triplet), without access to any other quantum numbers such as the $m_{ab}=-1,0,+1$ projection of the triplet~\cite{Ono2002,Johnson2005}. 
Already with two spins ($L=1$) we see that the four possible spin-states yield only one bit of measurement. 
For larger arrays, each pair returns one classical bit; $L$ pairs produce an $L$-bit string.
This coarse-grained readout distinguishes at most $2^L$ of the $\sim\!4^L / L^{3/2}$ basis states---a vanishing fraction as $L$ grows. 
This vanishing fraction is analogous to constructions of quantum error correction, in which an exponentially small set of measurements (e.g. corresponding to generators of the stabilizer group) are relied upon to correct and then measure against an exponentially larger set of states{.} 
However, QEC is constructed to explicitly limit the computational Hilbert space to the subspace of logical qubits, whereas our interest here is to utilize as much of the S0 space for computation as possible.

In this paper, we show that the utility of the S0 manifold for computation and its ability to demonstrate quantum advantage can be reliably benchmarked despite the vanishingly small fraction of measurements. 
We do this by adapting cross-entropy benchmarking (XEB), mirrored randomized benchmarking (MRB), and out-of-time-ordered correlators (OTOCs) to the S0 manifold with a coarse-grained PSB readout.

\begin{figure}[t]
	\includegraphics[width=0.8\columnwidth]{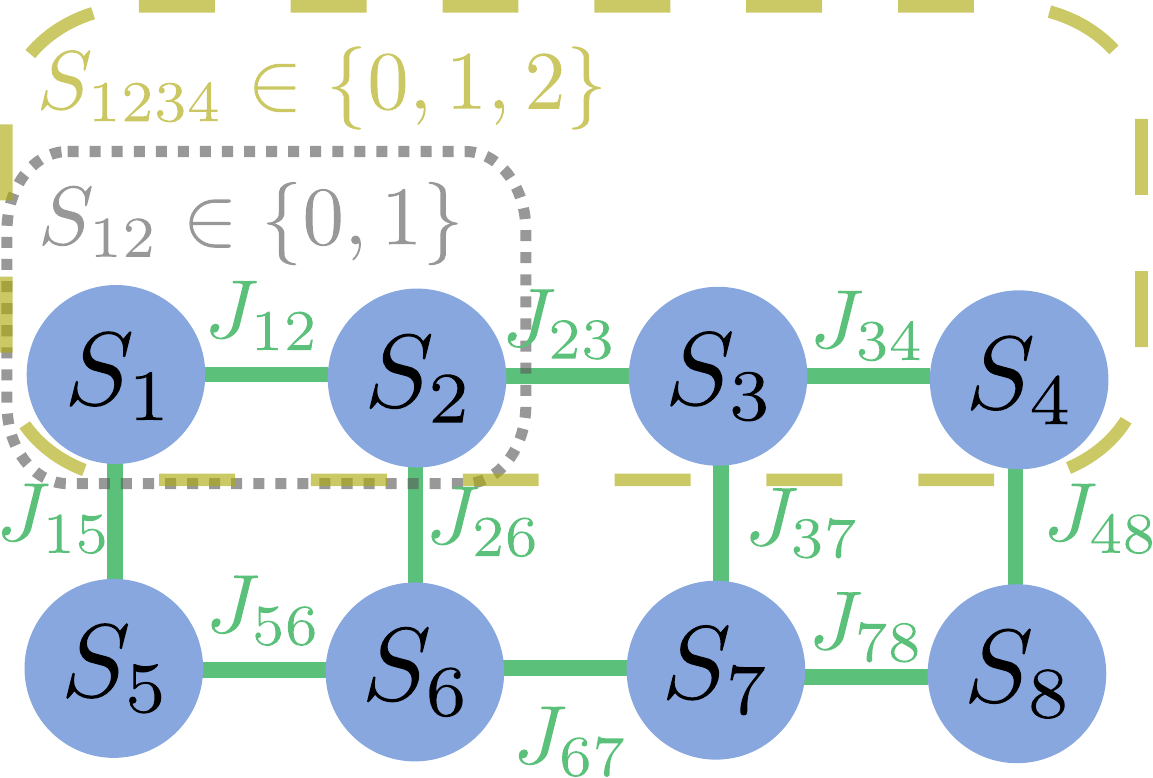}
	\caption{Layout of a $2\times4$ spin array and coupling with quantum numbers labeling the spin-zero manifold.}
	\label{fig:setup}
\end{figure}

In what follows, we will define the needed coarse-graining of PSB readout on the S0 system of $L$ spin-pairs, and use this limited measurement model to analyze and simulate the results of XEB, MRB, and OTOC experiments. 
Our main results, useful for the analysis of future experiments examining this system, are: 
\begin{enumerate}
\item The output probability distribution of random exchange circuits, coarse-grained by PSB, follows an Erlang distribution whose shape parameter equals the measurement-sector multiplicity, replacing the Porter--Thomas distribution familiar from standard qubit XEB.
\item
The coarse-grained linear XEB fidelity remains a faithful estimator of circuit quality.
\item
Under a Gaussian charge-noise model~\cite{Reed2016}, XEB fidelity decays predictably with circuit depth and noise strength, yielding a practical calibration tool that extracts average exchange error per pulse.
\item
OTOCs adapted to the S0 manifold can be used to generate large-loop interference and probe frustrated spins systems, opening a route to quantum advantage demonstrations in near-term spin arrays.
\end{enumerate}

These tools directly inform the gate fidelities governing fault-tolerant operations in exchange-coupled architectures~\cite{HRL2qubit2022,HRLmultirail2025}.

\begin{figure}
	\centering
	\begin{tikzpicture}[scale = 1.0,baseline=-10mm,level distance=10mm,level/.style={sibling distance=30mm/#1}]
		\node {$S_{12345678}=0$}
		child {
			node {$S_{1234}$} 
			child {node {$S_{12}$}
				child {node {$\frac{1}{2}$}}
				child {node {$\frac{1}{2}$}}}
			child {node {$S_{34}$}
				child {node {$\frac{1}{2}$}}
				child {node {$\frac{1}{2}$}}}}
		child {
			node {$S_{1234}=S_{5678}$}
			child {node {$S_{56}$}
				child {node {$\frac{1}{2}$}}
				child {node {$\frac{1}{2}$}}}			
			child {node {$S_{78}$}
				child {node {$\frac{1}{2}$}}
				child {node {$\frac{1}{2}$}}}};
	\end{tikzpicture}
	\caption{``Fusion tree" diagram for constructing the basis states of (in this case) $N=8$ spins, such as those in the $2\times 4$ array shown in Fig.~\ref{fig:setup}.  The bottom row shows $1/2$ for each individual spin, which are subsequently paired into $S_{12}$, $S_{34}$, etc.  These may each have value 0 (singlet) or 1 (triplet).  These pair into $S_{1234}=0,1$ or 2 which, for a total of 8-spins, must have equivalent value to $S_{5678}$ in order for the total spin $S=S_{12345678}$ to be zero.}
	\label{treefig}
\end{figure}
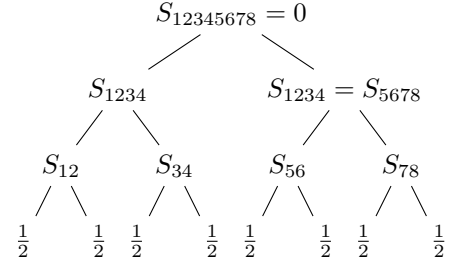

\section{Spin-zero manifold essentials.} 
We consider an array of semiconducting quantum dots, each hosting one electron.  (It is possible that a dot may additionally host closed-shells of electrons, with no significant impact on what follows.) 
Adjacent pairs of dots are initialized by relaxing two electrons in a charge-region of very high exchange, leaving them in the singlet state 
\begin{equation}
\ket{\Psi^-_i}=\frac{\ket{\uparrow_{2i-1}\downarrow_{2i}}-\ket{\downarrow_{2i-1}\uparrow_{2i}}}{\sqrt{2}}, 
\end{equation}
in terms of single-spin projection $S^z_i$ eigenstates \mbox{$\ket{\uparrow_i},\ket{\downarrow_i}$}.
Repeating this across a full device produces the total singlet
\begin{equation}
\ket{\psi_0}=\bigotimes_{i=1}^L\ket{\Psi^-_i}.
\end{equation}
The ``$z$" direction of this projection is arbitrary, as the singlet Bell state $\ket{\Psi^-_i}$ is itself an S0 manifold invariant to rotation.
In particular, no external magnetic field is needed to determine a projection axis, although when one is present, its direction is typically assigned to $z$.
Let $\vec{S}\cdot\vec{S}=S(S+1)=\sum_{i,j=1}^{2L}\vec{S}_i\cdot \vec{S}_j$ be the total spin operator of the spin system and  the vector of spin-1/2 operators.
The total spin of this initial state is zero, $\vec{S}^2\ket{\psi_0}=0$ (Fig.~\ref{fig:setup}).

Gate voltages electrically control the interaction between any two adjacent spins, which takes the form of a Heisenberg interaction, $H_{ij}=J_{ij}\vec{S}_i\cdot\vec{S}_j$, where $J_{ij}$ is the interaction strength between spins $i$ and $j$.
The rotational invariance of $H_{ij}$ assures $[\vec{S},H_{ij}]=\vec{0}$ for all $i,j$.  

The fusion-tree construction provides an orthonormal basis for the S0 subspace, characterizing states according to eigenvalues of the total spin of groups of $P$ spins, $S_{i_1\cdots i_P}$.
Such a tree is illustrated in Fig.~\ref{treefig}.
For the set of $L$ disjoint consecutive pairs of spins, one first assigns each pair a total spin $S_{2i-1,2i} \in \{0,1\}$, corresponding to singlet or triplet, respectively. 
Then one successively couples those pairs according to Clebsch-Gordan coefficients, assigning a quantum number corresponding to the total spin of each disjoint set of four spins, $S_{ijkl}\in\{0,1,2\}$. 
This process continues, iteratively coupling larger groups of spins until the coupling constrains all spins to total spin zero. 
For instance, for the $2\times4$ array considered in Fig.~\ref{treefig}, the basis states of the singlet subspace are eigenstates of $\vec{S}_{2i-1}\cdot\vec{S}_{2i}$ for $i\in\{1,2,3,4\}$ and $\vec{S}_{1234}^2$ with $\vec{S}_{1234}=\sum_{i=1}^4\vec{S}_i$. 

\begin{figure}[t]
	\includegraphics[width=1\columnwidth]{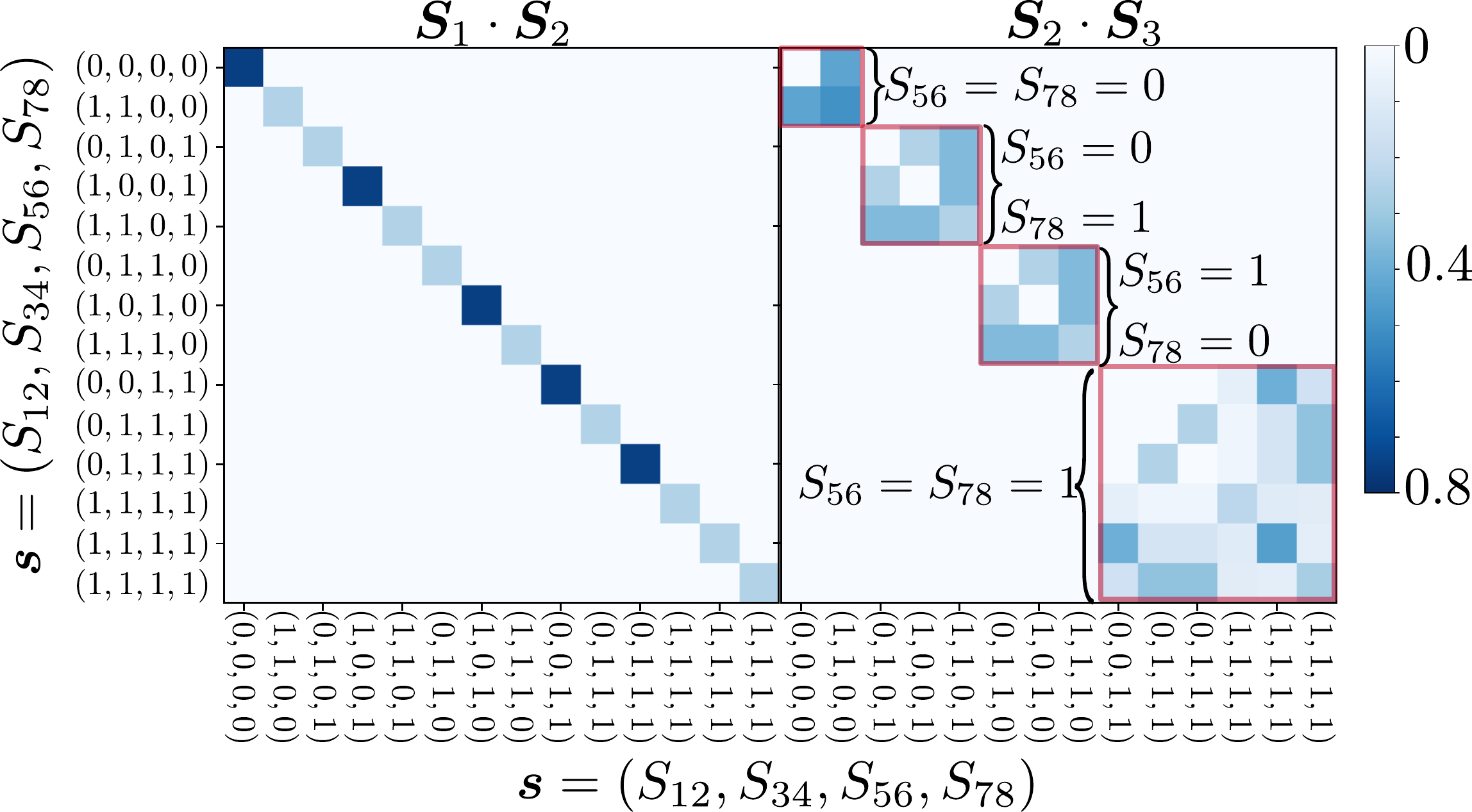}
	\caption{Magnitude of the elements of the unitary matrices $\bm S_1\cdot\bm S_2$ (left) and $\bm S_2\cdot\bm S_3$ (right) in the fusion-tree basis of a $2\times4$ array of spins.}
	\label{fig:ops}
\end{figure}

PSB readout on the $L$ pairs yields an $L$-bit string of singlet/triplet outcomes $\vec{s} = (S_{12},\ldots,S_{2L-1,2L})$.
These would be reflected by the  second-from-bottom layer of the fusion-tree example of Fig.~\ref{treefig}.
PSB does not resolve the three triplet magnetic substates of any pair, so it cannot distinguish basis states that share the same bitstring $\vec{s}$ but differ in higher-spin quantum number $S_{i_1\ldots i_P}$ for $P>2$ (i.e. higher layers of Fig.~\ref{treefig}). 
This motivates a convenient labeling of fusion-tree basis states, $\ket{\vec{s},\alpha}$, with $\vec{s}$ the measurement outcome string, and $\alpha$ an index for the states within the same measurement sector.
The multiplicity of outcome $\vec{s}$ is $R_{\vec{s}} = |\{\alpha\}|$ and equals the $w$-th Riordan number, with $w$ the Hamming weight of $\vec{s}$. 

In the 8-spin example of Fig.~\ref{treefig}, there would be $2^4=16$ possible singlet/triplet pair strings $\vec{s}$ across all $S=0,1,2,3,4$ manifolds.
The S0 manifold, however, admits only twelve bitstrings $\vec{s}$ because strings with exactly one triplet lie in $S=1$.
In this example, only the all-triplet outcome, $\vec{s}=(1,1,1,1)$, has nontrivial multiplicity, $R_{(1,1,1,1)} = 3$, and $\alpha$ indexes states with different eigenvalues of $S_{1234}$, i.e. 0, 1, or 2.
The total number of states in S0 is therefore $11+3=14=C_4$.  

Pulsing the Heisenberg exchange $H_{ij}$ for time $t$ generates the unitary $V_{ij}(\theta) = \exp(-i\theta\,\mathbf{S}_i{\cdot}\mathbf{S}_j)$, where $\theta = J_{ij}t$. In the fusion-tree basis, each $V_{ij}(\theta)$ is {represented as a $C_L\times C_L$ unitary matrix}.  
For a consecutive pair appearing in the fusion-tree basis, $V_{2i-1,2i}$ is diagonal with matrix elements
\begin{equation}
	\bra{\vec{s}',\alpha'}V_{2i-1,2i}\ket{\vec{s},\alpha}
	 = \delta_{\vec{s}'\vec{s}} \delta_{\alpha',\alpha} e^{3i\theta/4}e^{-i\theta S_{2i-1,2i}}\,,
\end{equation}
i.e. it acts as a phase gate on the singlet/triplet degree of freedom of that pair.
Otherwise, $V_{ij}$ mixes basis states and is block diagonal, where the conserved pair-spin quantum numbers, i.e., $S_{2k-1,2k}$ for $2k-1,2k\neq i,j$, determine the block structure (Fig.~\ref{fig:ops}).
Although $V_{ij}$ enables entangling gates between spins, there is no notion of single- and two-qubit gates within the S0 manifold.
Rather, exchange operations allow traversal throughout the reduced Hilbert space, analogous to the operation of $C_L$-dimensional qudit.

\begin{figure}[t]
	\includegraphics[width=0.9\columnwidth]{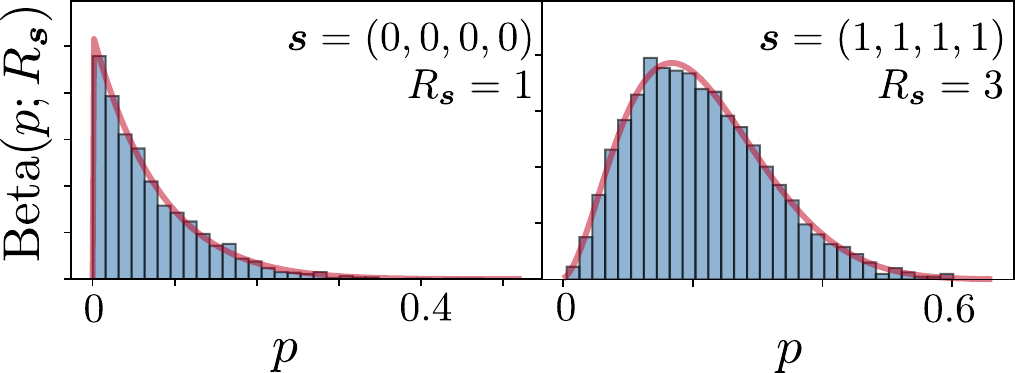}
	\caption{Simulated probability distributions of the all-singlet readout (left panel) and all-triplet readout (right panel) of a $2\times4$ array using 2000 circuits each with 100 randomly chosen exchange gates.Red envelopes show the exact Beta distribution.}
	\label{fig:dist}
\end{figure}

\section{Coarse-grained statistics} 
For benchmarking and quantum advantage experiments using the S0 sector of states $\ket{\vec{s},\alpha}$, we must work with the constraint of coarse-grained measurements that ignore $\alpha$.
Let $\ket{\psi_U}=U\ket{\psi_0}$ be a state in the S0 manifold, whose dimension we will now notate as $\Omega=C_L$, prepared by {an exchange-only} circuit corresponding to a unitary, $U=\prod_{m}V_{i_m j_m}(\theta_m)$.
We coarse-grain the probability of measuring a bitstring $\vec{s}$ over states with identical pair label:
\begin{equation}
  p_{\vec{s},U}
    = \sum_{\alpha=1}^{R_{\vec{s}}}
      \bigl|\braket{\vec{s},\alpha|\psi_U}\bigr|^2\,.
  \label{eq:coarse_prob}
\end{equation}
When $R_{\vec{s}}=1$, i.e., no degeneracy, $p_{\vec{s}}$ is a single Born-rule probability; when $R_{\vec{s}}>1$, it is a sum of $R_{\vec{s}}$ such terms.

\begin{figure}[t]
	\includegraphics[width=0.9\columnwidth]{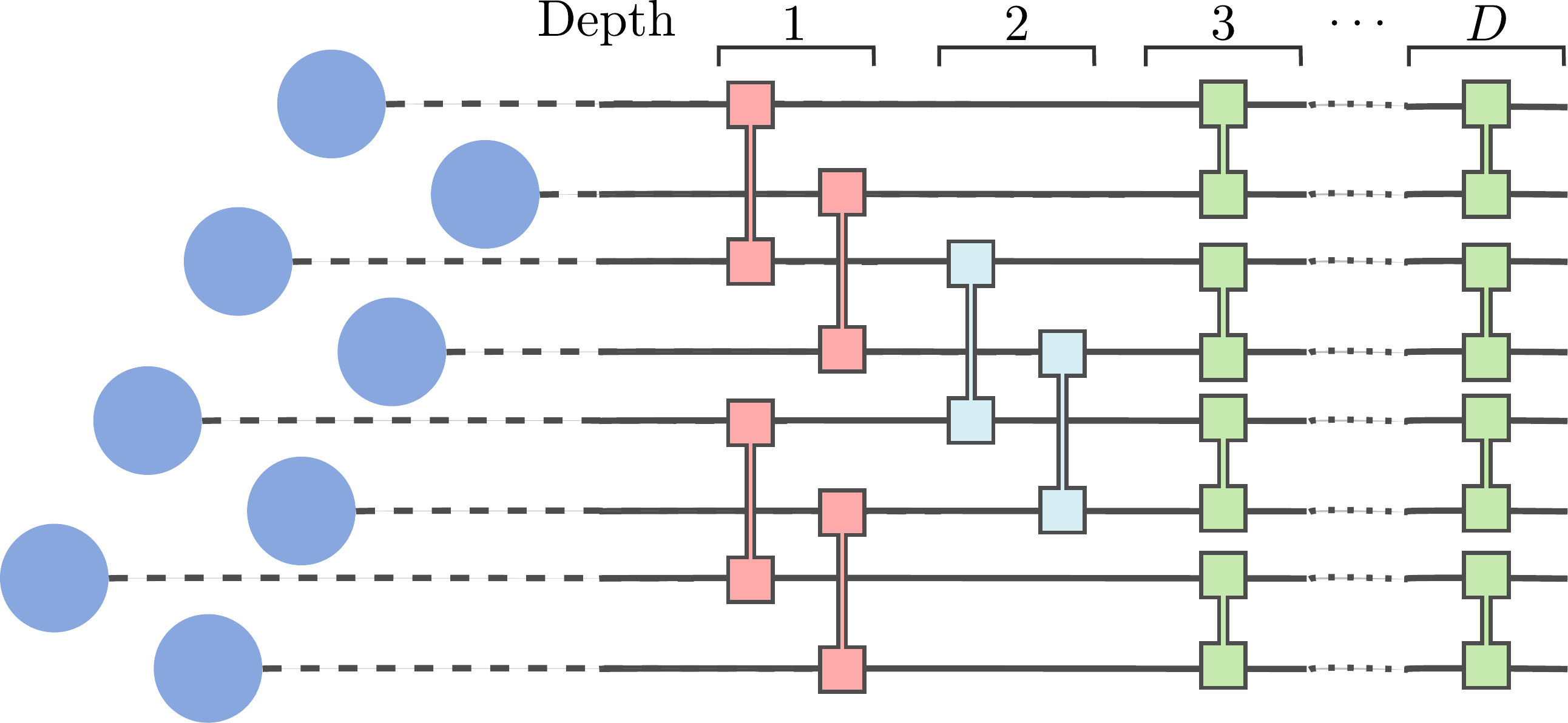}
	\caption{Sketch of the brickwork circuit of depth $D$ used to generate a random unitary matrix. Each two-spin interaction is an exchange gate $V_{ij}(\theta)$ with $\theta$ randomly drawn from a set of fixed angles.}
	\label{fig:circ}
\end{figure}

For benchmarking specifically, we also require a way to scramble the S0 sector. An exact 2-design for S0 is difficult to implement; we focus instead on random exchange circuits of depth $D$, consisting of $D$ brickwork layers of nearest-neighbor exchange gates with each angle drawn from $\theta\in\{\pi/5,2\pi/5,3\pi/5,4\pi/5,\pi/\sqrt{2}\}$ (Fig.~\ref{fig:circ}).
At sufficient depth, the resulting unitary $U_\textrm{Haar}$ approximates a Haar-random element of $\mathrm{U}(\Omega)$ within the S0 sector.
This set of gate angles is one of many that suffice to approximate a Haar-random circuit; a smaller number of angles could be used, or one could sample the angles from a uniform distribution, $\theta\in(0,\pi]$, and also obtain Haar-randomness.

A Haar-random state,  $\ket{\psi_\textrm{Haar}}=U_\textrm{Haar}\ket{\psi_0}$, can be realized by normalizing an i.i.d. complex Gaussian vector.
In the $\{\ket{\vec{s}',\alpha'}\}$ basis, this is given by
\begin{equation}
    \ket{\psi_\textrm{Haar}}
    =
    \frac{
    \sum_{\vec{s}',\alpha'} z_{\vec{s}',\alpha'}
    \ket{\vec{s}',\alpha'}
    }{
    \left(\sum_{\vec{s}',\alpha'} |z_{\vec{s}',\alpha'}|^2\right)^{1/2}
    }\,
\end{equation}
where the $z_{\vec{s}',\alpha'}$ are independent complex Gaussian random
variables, so that
$|z_{\vec{s}',\alpha'}|^2\sim \operatorname{Exp}(1)$. The probability of
obtaining the coarse-grained outcome $\vec{s}$ is therefore
\begin{equation}
    p_{\vec{s}} =
    \sum_{\alpha=1}^{R_{\vec{s}}}
    \bigl|\braket{\vec{s},\alpha | \psi_\textrm{Haar}}\bigr|^2
    =
    \frac{
    \sum_{\alpha=1}^{R_{\vec{s}}} |z_{\vec{s},\alpha}|^2
    }{
    \sum_{\vec{s}',\alpha'} |z_{\vec{s}',\alpha'}|^2
	}\,.
\end{equation}
The numerator $Z_\vec{s}\equiv \sum_{\alpha=1}^{R_\vec{s}}|z_{\vec{s},\alpha}|^2$ is a sum of $R_{\vec{s}}$ independent exponential random variables, so it follows a gamma distribution $Z_\vec{s}\sim \Gamma(R_{\vec{s}},1)$.
The denominator contains every term in the numerator, plus $\Omega - R_{\vec{s}}$ more random variables that follow an independent distribution $Z_{s'}\equiv \sum_{s'\neq s, \alpha'}|z_{\vec{s}',\alpha}|^2 \sim \Gamma(\Omega - R_{\vec{s}},1)$.
Expressing $p_\vec{s}$ in terms of these gamma-distributed variables, we see that it follows a beta distribution (Fig.~\ref{fig:dist}),
\begin{equation}
    p_\vec{s} = \frac{Z_\vec{s}}{Z_\vec{s} + Z_{s'}} \sim \textrm{Beta}(R_{\vec{s}},\Omega-R_{\vec{s}})\,,
\end{equation}
with PDF
\begin{equation}
    f(p_\vec{s};\,R_\vec{s}, \Omega) =R_\vec{s}\binom{\Omega-1}{R_{\vec{s}}}p_{\vec{s}}^{R_\vec{s}-1}(1-p_{\vec{s}})^{\Omega-R_{\vec{s}}-1}\,.
\end{equation}
The beta distribution has first and second moments $\langle p_\vec{s} \rangle=R_\vec{s}/\Omega$ and $\langle p_\vec{s}^2 \rangle=R_\vec{s}(R_\vec{s}+1)/[\Omega(\Omega+1)]$, respectively. 

In the large-$\Omega$ limit, the beta distribution becomes the Erlang distribution, 
\begin{equation}
  f(p_\vec{s};\,\,R_\vec{s},\Omega) = \frac{\Omega^{R_\vec{s}}\, p_\vec{s}^s{R_\vec{s}-1}}{(R_\vec{s}{-}1)!}\,e^{-\Omega p_\vec{s}}\,,
  \label{eq:erlang}
\end{equation}
with $\langle p_\vec{s}\rangle =R_\vec{s}/\Omega$ and $\langle p_\vec{s}^2\rangle =R_\vec{s}^2/\Omega^2$. 
When the multiplicity is one, we recover the Porter--Thomas distribution, $f(p;\,\Omega)=\Omega\exp(-\Omega p)$, which is the typical distribution of measurement probabilities of Haar-randomized quantum states with non-degenerate measurement. 

\section{Benchmarking}
We can use randomized circuits to benchmark the quality of the exchange-only dynamics.
Randomized benchmarking is impractical, because unlike Clifford gates, random sequences of exchange gates cannot be efficiently inverted.
Cross-entropy benchmarking (XEB) and mirror randomized benchmarking (MRB), by contrast, can be adapted to S0.

Beginning with XEB, for a random circuit $U$ and bitstring $\vec{s}$, let $p^m_{\vec{s},U}$ denote the experimentally observed probability to measure $\vec{s}$.
We parameterize $p^m_{\vec{s},U}$ as a linear combination of the ideal probability $p_{\vec{s},U}$ and some other distribution $\chi_{\vec{s},U}$:
\begin{equation}
  p^m_{\vec{s},U} = F p_{\vec{s},U} + (1-F) \chi_{\vec{s},U}.
\end{equation}

Following Ref.~\cite{boixoNATP18}, we could obtain the XEB formula by assuming a depolarizing noise model and performing maximum-likelihood estimation of $F$.
But for exchange-only spins, this is a dubious assumption.
Magnetic gradients cause leakakge out of the S0 manifold, which alters the measurement degeneracies $R_\vec{s}$ and causes $\chi$ to deviate from the depolarized distribution even under perfect Haar-averaging.
Instead, we assume that $\chi$ may be a combination of: (1) the depolarized distribution, (2) an incorrect Haar-random distribution, resulting from $U$-dependent but coherent error, (3) a leaked distribution, (4) $U$-independent SPAM error, and (5) fluctuations due to finite statistics.

The magnitudes of each component may not be known in advance, but under sufficiently scrambling random circuits, we expect all of them to be statistically independent of $p_{\vec{s},U}$, i.e. $\operatorname{Cov}_U(p_{\vec{s},U}, \chi_{\vec{s},U})=0$.
This implies
\begin{equation}
  F \propto \operatorname{Cov}_U(p^m_{\vec{s},U}, p_{\vec{s},U}).
\end{equation}
Moreover, all three types of error channels (depolarizing, Haar-randomizing, and leakage) reduce state fidelity to $O\left(\Omega^{-1}\right)$, so in the large-$\Omega$ limit $F$ can be interpreted as a SPAM-adjusted fidelity estimate.
Thus, the normalized covariance of $p^m$ with $p$ provides an independent estimate of $F$ for each $\vec{s}$.
The overall estimate's signal-to-noise ratio can be improved by normalizing each term with the ideal expected probability $\langle p_\vec{s}\rangle$, leaving
\begin{align}\label{eq:xeb}
F \approx \mathcal{N}&\left(\sum_\vec{s}\frac{\langle p^m_\vec{s} p_\vec{s} \rangle}{\langle p_\vec{s}\rangle} - 1\right)\,,\\
\mathcal{N} \equiv &\left(\sum_\vec{s}\frac{\langle p_\vec{s}^2 \rangle}{\langle p_\vec{s}\rangle} - 1\right)^{-1}\,.\nonumber
\end{align}
When measurement is non-degenerate and in the limit of many random circuits, $\langle p_\vec{s}\rangle^{-1}=\Omega$, and this reduces to the standard formula for linear XEB.

Note that multiplicity increases the statistical cost of resolving XEB.
As the number of spins grows, the highest-weight outcomes correspond to a growing number of unresolved fine-grained fusion states.
A measured probability in such a sector is a sum over many fine-grained Haar probabilities, so by the central limit theorem it is more concentrated around its mean value and has less relative Porter--Thomas contrast.
This reduces the distinction between ideal sampling and reference sampling in the coarse-grained distribution.
For depolarizing or coherent exchange noise, the variance of the coarse-grained XEB signal scales as $C_L/4^L$, which still decreases with increasing $L$. But noise channels that bias the outcome toward high-weight bitstrings, such as SPAM or leakage errors, may require averaging over an increasing number of random circuits with system size.
XEB is also limited by the need to simulate $p_\vec{s}$, so even without coarse-grained measurement we are limited to classically simulable system sizes.

These caveats motivate using MRB as an additional benchmark.
In MRB, we estimate fidelity by measuring the return probability upon application of a random unitary and its inverse,
\begin{equation}
    F_\textrm{MRB}
      =
      \left\langle
      \left|
      \langle\psi_0|U^{-1}U|\psi_0\rangle
      \right|^2
      \right\rangle_U\,.
    \label{eq:MRB}
\end{equation}
We invert each exchange gate $V_{i j}(\theta)$ with a complementary gate $V_{i j}(2\pi-\theta)$, and we invert the full $U$ by applying the inverted gates in reverse time order, i.e. $U^{-1}=\prod_{m=\textrm{Max}(m)}^0 V_{i_m j_m}(2\pi-\theta_m)$.
An ideal MRB circuit will always return $\vec{s}=(0,0,...0)$, so no simulation is needed.
Moreover, the all-singlets outcome is not degenerate, so coarse-graining does not affect the analysis.
These make MRB a useful complementary benchmark to XEB, especially for larger system sizes.

We simulate XEB in sufficiently small arrays by applying noisy random unitary circuits of depth $D$ to $\ket{\psi_0}$, from which we extract $p^m_{\vec{s},U}$.
We also apply the corresponding noiseless circuit to obtain $p_{\vec{s},U}$, from which we compute XEB fidelity according to Eq.~(\ref{eq:xeb}).
For comparison, we also calculate a fine-grained XEB fidelity by resolving
$p_{\vec{s}\alpha}=|\langle \vec{s},\alpha|\psi_U\rangle|^2$
and
$p^m_{\vec{s}\alpha}=|\langle \vec{s},\alpha|\psi^m_U\rangle|^2$.
We find that the (experimentally inaccessible) fine-grained benchmark gives nearly identical fidelities to its coarse-grained counterpart, confirming that the results are robust to coarse-graining.

To simulate MRB, we execute the inverse of each circuit,inverting each $\theta$ exchange with a $2\pi-\theta$ exchange as described above; we denote this $2\pi$-MRB. 
As a point of comparison, we also apply MRB by explicitly taking $V_{ij}(\theta)\rightarrow V_{ij}(-\theta)$, which we call negative MRB (NMRB).
NMRB is not experimentally accessible, but is more directly comparable with XEB, since the original and inverting gate pulses have the same magnitudes.

\begin{figure}[t]
	\includegraphics[width=\columnwidth]{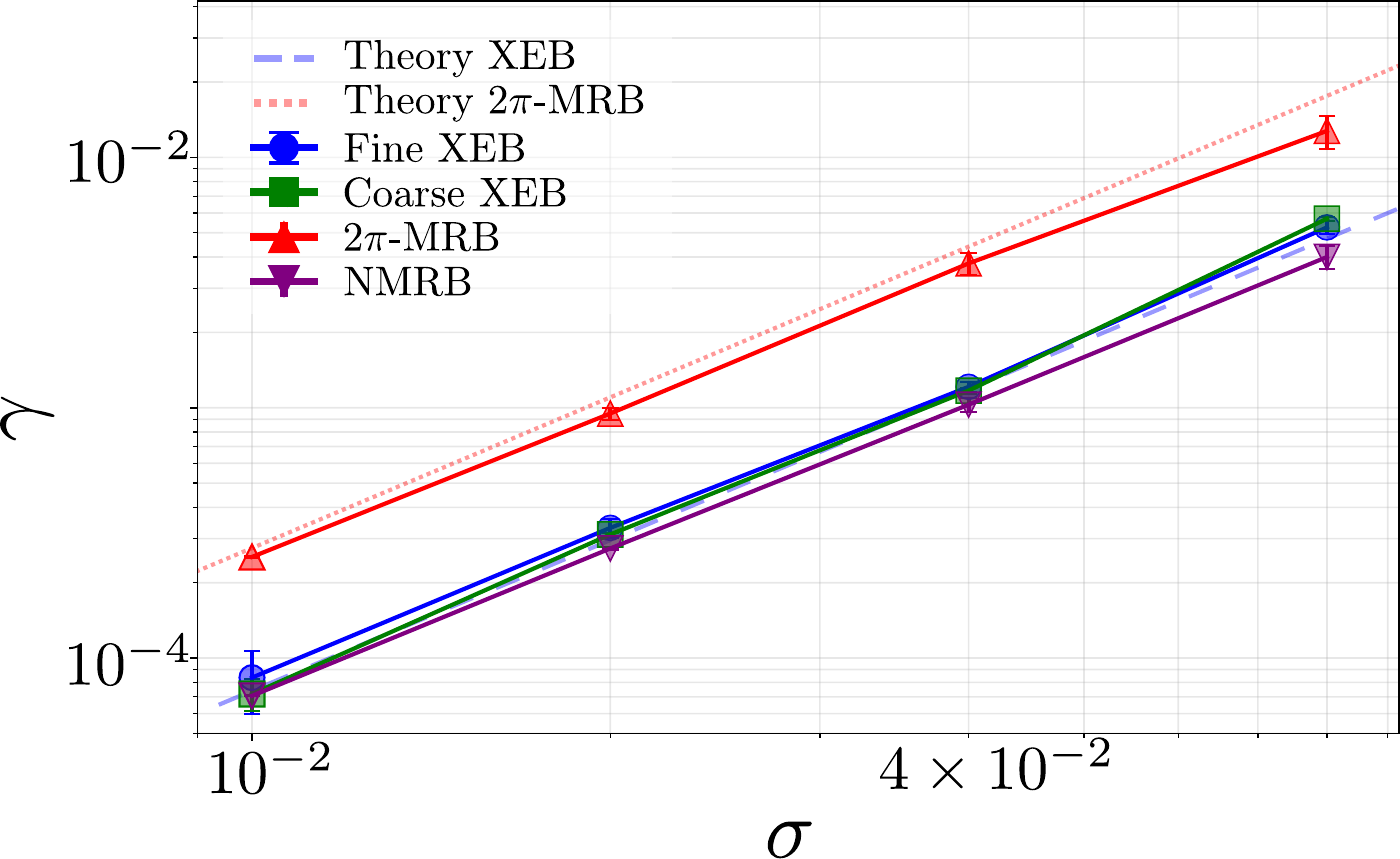}
	\caption{Average error per gate, $\gamma$, as a function of charge noise variance, $\sigma$, extracted from numerical simulations using fine XEB (circles), coarse XEB (squares), $2\pi$-MRB (up-triangles), and NMRB (down-triangles). Theoretical prediction of $\gamma$ for fine XEB, coarse XEB, and NMRB (dashed curve) and $2\pi$-MRB (dotted curve).}
	\label{fig:benchmark}
\end{figure}

We focus on benchmarking charge noise, which we model as fluctuations in the exchange interaction, $J_{ij}$. 
For an ideal unitary $V_{ij}(\theta)$, the noisy analogue is $V_{ij}(\theta+\epsilon\theta)$ where we draw $\epsilon$ from a normal distribution with zero mean and variance $\sigma$.
The expected fidelity is $F=\exp(-\gamma D)$, $\gamma=\left\{\langle\theta^2\rangle\textrm{Tr}[(\bm{S}_i\cdot\bm{S}_j)^2]/\Omega\right\}\sigma^2$. On a $2\times4$ array of spins, for a given value of $\sigma$, over $1000$ random circuits, we calculate the fidelity as a function of circuit depth $D$ using coarse XEB, fine XEB, $2\pi$-MRB, and NMRB and extract $\gamma$. 
In Fig.~\ref{fig:benchmark}, we plot $\gamma$ as a function of $\sigma$ and find that the error per gate that we extract via XEB and NMRB agrees well with the theoretical prediction.
While $2\pi$-MRB yields a characteristically larger error per gate because the average sampled angle is larger, it also matches the theoretical predictions well.

\section{Quantum supremacy} 
Quantum supremacy refers to the out-performance of a quantum computer over any classical one for some task, regardless of the utility of that task.
Although XEB can serve as a metric for quantum supremacy~\cite{boixoNATP18}, with S0 the aforementioned contrast of coarse-grained XEB scales poorly with the system size, requiring a large number of random circuits to estimate XEB accurately, and significantly weakening XEB-based supremacy. 
Analogously to using MRB for benchmarking, one can bypass coarse-graining entirely by focusing on echo-based circuits to demonstrate supremacy. 
A well-known example is the out-of-time-ordered correlator of order $k$ (OTOC$^k$), which we define as
\begin{equation}
    \mathcal{C}^{(2k)}=\langle\psi_0|(U^\dagger B U M)^{2k}|\psi_0\rangle\,,
    \label{eq:OTOCdef}
\end{equation}
where $M$ is a Hermitian measurement operator, $B$ is the so-called butterfly operator, and $U$ describes the {time}-evolution of $B$ 
\footnote{When $B=I$ and $M=|\psi_0\rangle\langle\psi_0|$, we obtain the circuit for MRB, $\mathcal{C}^{(2)}=\langle\psi_0|(U_\textrm{Haar}^\dagger U_\textrm{Haar} |\psi_0\rangle\langle\psi_0| )^{2}|\psi_0\rangle=|\langle\psi_0|(U_\textrm{Haar}^\dagger U_\textrm{Haar} |\psi_0\rangle|^2$.}.  
For the moment, let us focus on evolution according to a random circuit, $U=U_\textrm{Haar}$. 
Analogous to strings of Pauli operators in a conventional qubit architecture, we use Weyl--Heisenberg operators, $W^{\mu}\equiv W^{a,b}=X^a Z^b$ with $\mu$ as a superindex over pairs of $a,b\in\{0,\ldots,\Omega-1\}$, to provide a complete orthonormal basis, i.e., $\textrm{Tr}[W^\mu W^\nu]=\delta_{\mu\nu}$, for any $\Omega$-dimensional unitary matrix. 
Here, $X=|j\rangle\langle j+1|$, the shift operator, and $Z=\omega^j|j\rangle\langle j|$, the clock operator, obey the relation $XZ=\omega ZX$ with $\omega=\exp(2\pi i/\Omega)$. 
In the ergodic limit, decomposing the evolved butterfly operator as {$U^\dagger B U=\sum_{\mu=1}^{\Omega^2} b_\mu W^\mu$},
\begin{align}
	\mathcal C^{(2)}&=\sum_{\mu\nu}c_{\mu\nu}^{(2)}\textrm{Tr}[W^\mu W^{\nu} ]/\Omega\,,\nonumber\\
	\mathcal C^{(4)}&=\sum_{\mu\nu\lambda\kappa}c^{(4)}_{\mu\nu\lambda\kappa}\textrm{Tr}[W^\mu W^\nu W^{\lambda} W^{\kappa}]/\Omega\,,
	\label{eq:OTOCtr}
\end{align}
where $c^{(2)}_{\mu\nu}=b_\mu b_{\rho} T^M_{\rho\nu}$, $c^{(4)}_{\mu\nu\lambda\kappa}=b_\mu b_{\rho}b_\lambda b_\sigma T^M_{\rho\nu}T^M_{\sigma\kappa}$, and $T^M_{\nu\lambda}=\textrm{Tr}[W^\lambda M W^\nu M]$ is the Weyl--Heisenberg transfer matrix for the operator $M$. 
In contrast to the usual qubit case, where $M$ is a Pauli string \cite{googleqcNAT25}, in the spin-zero manifold $M$ is typically a PSB measurement on adjacent spins, which in general decomposes into a sum of multiple Weyl--Heisenberg operators in the fusion-tree basis.  
For $L$ even, choosing any set of $L-1$ PSB measurements of consecutive pairs appearing in the fusion-tree basis yields $M=Z^{\Omega/2}$ up to a reordering of the S0 basis \footnote{Using the Murnaghan-Nakayama rule, the character of $L-1$ disjoint two-cycles in the singlet representation of $L$ spins is zero, i.e. the singlet representation of such an element of the permutation group is traceless. Choosing disjoint consecutive pairs, this representation must be diagonal with entries of $\pm1$. Consequently, the matrix representation must be $Z^{\Omega/2}$ up to a reordering of the fusion-tree states.}, e.g., for $L=4$, $M=U_{12}(\pi)U_{34}(\pi)U_{56}(\pi)\approx Z^{7}$, rendering the transfer matrix diagonal. 

Similar to the conventional qubit case, a phase that depends on $M$ and the evolution details dresses each element in the summations in Eq.~(\ref{eq:OTOCtr}), but the magnitude stays the same. 
We can now leverage arguments analogous to those made in Ref.~\onlinecite{googleqcNAT25}: while orthogonality of Weyl--Heisenberg operators demands, in $\mathcal C^{(2)}$, that $W^\mu=(W^\nu)^\dagger$, the trace in $\mathcal C^{(4)}$ is nonzero when $W^\mu W^\nu W^{\lambda} W^{\kappa}=\boldsymbol{1}$, which allows for many large-loop interference paths. 
A Monte Carlo simulation can calculate $\mathcal C^{(2)}$ relatively easily, while $\mathcal C^{(4)}$ becomes difficult for a sufficiently large system size and circuit depth \cite{googleqcNAT25}.

\section{Quantum advantage} 
Quantum advantage refers to the use of the quantum computer to perform a useful task, offering a performance advantage relative to a classical computation. OTOCs can be adapted to probe classically inaccessible properties of physical systems, described by a Hamiltonian $\mathcal H$, by using the associated time-evolution operator, $U\rightarrow U_{\mathcal H}(t)=\exp(-i \mathcal H t/\hbar)$ in Eq.~(\ref{eq:OTOCdef}).
Specifically, because the exchange interaction is native to spin arrays, we focus on computing $\mathcal C_\mathcal{H}$ for  Hamiltonians of the form
\begin{equation}
	\mathcal H=\sum_{ij}J_{ij}\bm S_i\cdot \bm S_j\,.
	\label{eq:ham}
\end{equation}
Because we can emulate non-local interactions through judicious use of the native SWAP gates, $V_{ik}(\theta)=V_{ij}(\pi)V_{jk}(\theta)V_{ij}(\pi)$, the model supports a general lattice, in any dimension, with arbitrary exchange couplings. 
By computing $\mathcal C_\mathcal{H}$ at various times $t$ and spin locations $\{i,j,k,l\}$, one can extract quantities such as butterfly velocity, scrambling time, and front broadening exponent to characterize properties of the model such as the spectral gap and integrability.
\begin{widetext}
	
Although a detailed Trotter schedule to enable arbitrary $J_{ij}$ in Eq.~(\ref{eq:ham}) is beyond the scope of our current analysis~\cite{burkardSciPo25},
as one concrete example, we consider a spin chain with next-to-nearest-neighbor Heisenberg interactions, \hbox{$\mathcal H=\sum_{i=1}^{N-1} J_1 \bm S_i\cdot\bm S_{i+1} + \sum_{i=1}^{N-2}J_2 \bm S_i\cdot\bm S_{i+2}$}, whose rich phase structure is controlled by $\phi=J_2/J_1$ \cite{majumdarJoMP69a, majumdarJoMP69b, haldanePRB82, okamotoPLA92, whitePRB96}. To Trotterize, we break up the Hamiltonian into five parts consisting of sums of maximally commuting elements, $\mathcal{H}=\sum_{j=1}^5\mathcal{H}_j$ with
\begin{align}
	\mathcal H_1&=\sum_{i=1}^{\lfloor N/2\rfloor} {J_1} \bm S_{2i-1}\cdot\bm S_{2i}\,, \hspace{1cm} \mathcal H_2=\sum_{i=1}^{\lfloor (N/2) -1\rfloor} {J_1} \bm S_{2i}\cdot\bm S_{2i+1}\,,\nonumber\\
	\mathcal H_3&=\sum_{i=1}^{\lfloor N/3\rfloor} {J_2} \bm S_{3i-2}\cdot\bm S_{3i}=\left(\prod_{i=1}^{\lfloor N/3\rfloor}\textrm{SWAP}_{3i-2,3i-1}\right)\left(\sum_{i=1}^{\lfloor N/3\rfloor} {J_2} \bm S_{3i-1}\cdot\bm S_{3i}\right)\left(\prod_{i=1}^{\lfloor N/3\rfloor}\textrm{SWAP}_{3i-2,3i-1}\right)\,,\nonumber\\
	\mathcal H_4&=\sum_{i=1}^{\lfloor (N-1)/3\rfloor} {J_2} \bm S_{3i-1}\cdot\bm S_{3i+1}=\left(\prod_{i=1}^{\lfloor (N-1)/3 \rfloor}\textrm{SWAP}_{3i-1,3i}\right)\left(\sum_{i=1}^{\lfloor (N-1)/3 \rfloor} {J_2} \bm S_{3i}\cdot\bm S_{3i+1}\right)\left(\prod_{i=1}^{\lfloor (N-1)/3 \rfloor}\textrm{SWAP}_{3i-1,3i}\right)\,,\nonumber\\
	\mathcal H_5&=\sum_{i=1}^{\lfloor (N-2)/3\rfloor} {J_2} \bm S_{3i}\cdot\bm S_{3i+2}=\left(\prod_{i=1}^{\lfloor (N-2)/3 \rfloor}\textrm{SWAP}_{3i,3i+1}\right)\left(\sum_{i=1}^{\lfloor (N-2)/3 \rfloor} {J_2} \bm S_{3i+1}\cdot\bm S_{3i+2}\right)\left(\prod_{i=1}^{\lfloor (N-2)/3 \rfloor}\textrm{SWAP}_{3i,3i+1}\right)\,.
\end{align}
Decomposing the nonlocal terms according to $\mathcal H_{i\in[3,4,5]}\equiv \textrm{SW}_i\, \tilde{\mathcal H}_i\, \textrm{SW}_i$ such that $\tilde{\mathcal H}_i$ is the local exchange Hamiltonian and $\textrm{SW}_i$ is the product of SWAPs that emulates non-locality, the Trotterized Hamiltonian is
\begin{equation}
	U_\textrm{Trotter}(t)=
	\left(\prod_{j=1}^5\exp(-i \mathcal{H}_j \Delta t)\right)^{\mathcal{N}} =
	\left(
	e^{-i \mathcal{H}_1 \Delta t} e^{-i \mathcal{H}_2 \Delta t} 
	\textrm{SW}_3 e^{-i \tilde{\mathcal{H}}_3 \Delta t} {\textrm{SW}_3} 
	\textrm{SW}_4 e^{-i \tilde{\mathcal{H}}_4 \Delta t} \textrm{SW}_4 
	\textrm{SW}_5 e^{-i \tilde{\mathcal{H}}_5 \Delta t} \textrm{SW}_5
	\right)^\mathcal N\,,
\end{equation}
where the total time of evolution is $t=\mathcal N\Delta t$ and $\mathcal N$ is the number of Trotter steps. 
\end{widetext}
Using $U_{\mathcal H}=U_\textrm{Trotter}$, we expect the properties of the system to inform $\mathcal C_{H}$ \cite{xuPRXQ24}. 
For instance, when $\phi<0.24$ ($\phi>0.24$), the system is in a gapless Luttinger liquid phase (gapped valence bond solid) and we expect the butterfly velocity to be relatively large (small). 
Moreover, going from the commensurate phase, $0.24<\phi<0.5$, to the incommensurate phase, $\phi>0.5$, spiral order emerges, which should be evident in $\mathcal C_{H}$.

As the Trotter error for a fixed set of Hamiltonian parameters ($J_1,J_2$) is generally a function of $N\Delta t$, the step size $N$ must be reduced proportional to the size of the system to keep a Trotter approximation stable. Furthermore, as the number of pulses required to simulate a Hamiltonian for a fixed amount of time scales like $N/\Delta t$, the total number of pulses to simulate this Heisenberg Hamiltonian grows like $N^2$. This puts a strict constraint on the necessary gate fidelity. If one wants to simulate a system with ten-times more spins at a similar level of accuracy, the gate fidelity must improve one-hundred fold. Trotterization strategies for specific Hamiltonians can be explored to help soften this constraint, but note that higher order Trotter approximations likely come at the cost of a higher routing overhead, complicating total circuit error optimization \cite{zhangCM25}.

\section{Outlook} 
At time of writing, state-of-the-art semiconducting spin systems host a $3\times18$ array of spins, roughly equivalent to $47$ qubits when using the spin-zero manifold, operating at exchange fidelities nearing $99.99\%$ \cite{abrahamCM26}. 
Because superconducting systems have a comparable speed of operation to semiconducting spin systems, unlike neutral-atom or ion-trap qubit systems which have slower clock speeds, it is apt to compare the former with spins. State-of-the-art superconducting-based quantum computers host roughly twice as many qubits as spin systems today. However, average single- and two-qubit fidelities are $99.95\%$ and $99.85\%$, respectively \cite{googleqcNAT25}. 
Although {current} spin systems {do not} beat superconducting qubits in {the size of the usable Hilbert space (e.g. qubit count)}, we anticipate that their ability to run deep circuits {(albeit with depth measured in exchange operations rather than CNOT gates)} will support competitive claims of supremacy and advantage.  

While we have gone into detail about the {experimental} operation of the spin-zero manifold, we have given only three examples of how to use it. We anticipate additional applications to supremacy claims via heavy output generation and parameter estimation in Heisenberg spin systems with OTOC variants.
{And} although we restricted our analysis to the coarse-grained outputs, efficient readout of the full spin-zero state may also be achievable.
One could do this by iterative readout of different combinations of spin pairs to tomographically extract the circuit output. 
Alternatively, appropriately prepared ancilla spins could differentiate between triplet substates.

Our analysis of the spin-zero manifold extends naturally to other physical systems.
For instance, one could perform such an analysis on holes in germanium rather than spins in silicon. In that case, the system no longer conserves spin. Nonetheless, if one knows the unitary structure a priori, there may exist an analogous set of benchmarks and OTOCs. 

Rather than utilizing the full S0 manifold, one could modularize over groups of six or more spins. In the six spin example, one could effectively operate coupled qudits of dimension five, working in a subspace of the full S0 Hilbert space. This idea can be generalized to modules of uniformly large or inhomogeneous sizes, and combinations could reconcile the advantages of S0 operation with practical methods for fault-tolerant error correction.  

\begin{acknowledgments}
We thank Joe Kerckhoff and Dwight Luhman for helpful discussions. CT would like to thank ``the automotans''---Opus 4.6 and others in copilot CLI who helped ideation of this project over the holidays. SH, EHC, MB, and CT used copilot CLI for code development.
\end{acknowledgments}

\bibliographystyle{apsrev4-2}
\bibliography{references}

\end{document}